\documentclass[draft,jgrga]{AGUTeX}

\usepackage{slashbox}
\usepackage{multirow}
\usepackage{multicol}

\usepackage{color}
\newcommand{\red}{\textcolor{black}}

\usepackage{bm}

\usepackage{graphicx}
\graphicspath{{pics/},{pic/}}

\setkeys{Gin}{draft=false}

\usepackage{url}
\usepackage{bm}
\usepackage{amsmath}

\authorrunninghead{LULU ZHAO ET AL.}

\titlerunninghead{TRANSPORT OF ENERGETIC PARTICLES IN CIRs}

\authoraddr{Corresponding author: Gang Li, Department of Space Science and CSPAR, 
University of Alabama in Huntsville, Huntsville, Alabama, USA. (gang.li@uah.edu)}

\begin{document}

\title{MODELING TRANSPORT OF ENERGETIC PARTICLES IN COROTATING INTERACTION REGIONS -- A CASE STUDY}




\authors{Lulu Zhao \altaffilmark{1}, Gang Li\altaffilmark{1}, 
R. W. Ebert \altaffilmark{2}, M. A. Dayeh \altaffilmark{2}, 
M. I. Desai \altaffilmark{2}, G. M. Mason \altaffilmark{3}, 
Z. Wu \altaffilmark{4}, Y. Chen \altaffilmark{4} }

\altaffiltext{1}{Department of Space Science and CSPAR, University of Alabama in Huntsville, Huntsville, Alabama, USA.}

\altaffiltext{2}{Southwest Research Institute, San Antonio, TX, USA}

\altaffiltext{3}{Applied Physics Laboratory, Johns Hopkins University, Laurel, MD, USA}

\altaffiltext{4}{Institute of Space Sciences and School of Space Science and Physics, 
Shandong University at Weihai, Weihai, China}



\begin{abstract} We investigate energetic particle transport in Corotating Interaction Regions (CIRs) through a case study. The CIR event we study occurred on $2008$ February $08$ and was  observed by both the Advanced Composition Explorer (ACE) and the twin Solar TErrestrial RElations Observatory (STEREO)-B  spacecraft. An in-situ reverse shock was observed by STEREO-B ($1.0$ AU) but not ACE ($0.98$ AU). Using STEREO-B observations and assuming the CIR structure does not vary significantly in the corotating frame,  we estimate the shock location at later times for both the STEREO-B and ACE observations. Further assuming the accelerated particle spectral shape at the shock does not vary with shock location, we calculate the particle differential intensities as observed by ACE and STEREO-B at two different times  by solving the focused transport equation using a Monte-Carlo simulation.  We assume that particles move along Parker's field and experience no cross-field diffusion. We find that the modulation of sub-MeV/nucleon particles is significant. To obtain  reasonable comparisons between the simulations and the observations by both ACE and STEREO-B, one has to assume that  the CIR shock can accelerate more particles at a larger heliocentric distance than at a smaller heliocentric distance.  
\end{abstract}


\begin{article}

\section{Introduction} 
Corotating Interaction Regions (CIRs) are formed when fast solar wind, originating from coronal holes that  extend to low latitudes, compresses slow solar wind in front of it. A forward shock can form at the  leading edge of the compression region and propagate into the slow solar wind, and a reverse shock can form at  the trailing edge of the compression region and propagate into the fast solar wind.  Early studies showed that by $3$ to $4$ Astronomical Units (AU), most CIRs are bounded by a forward and  reverse shock pair \citep[e.g.][]{Hundhausen1976}. Recently, a statistical study by \citet{Jian2006} found  that $31\%$ of CIRs observed at 1 AU are associated with shocks.

 CIRs are a major source of energetic particles in the inner heliosphere during solar minimum  \citep[e.g.][]{VanHollebeke1978, McDonald1976, Richardson1993}.  \citet{Fisk1980} first examined particle acceleration associated with CIRs. They solved the steady state  transport equation with a geometry appropriate to CIRs. In their model, particles are accelerated via the first order Fermi acceleration mechanism at either the forward or the reverse shock which are often at a  distance of several AU. Energetic particles then propagate along the interplanetary magnetic field back to $1$ AU.  Similar to Galactic Cosmic Rays (GCRs) and Anomalous Cosmic Rays (ACRs), the adiabatic cooling can lead to a  modulation effect at low energies.

Later, \citet{Giacalone2002} proposed another mechanism which, instead of invoking shocks,  considered particle acceleration in gradual and slowly-varying solar wind compression regions. They  found that particles can be accelerated up to $10$ MeV/nucleon by a process similar to diffusive shock acceleration,  in which particles gain energy by scattering between converging scattering centers. Simulation results similar to observations were obtained for a reasonable set of parameters.

Recent observations by \citet{Mason1997} suggested that the observed energy spectra in CIR events often do not show modulation effects 
in low energies. Instead, the spectra continue to fall down as a power law from the energy threshold of 
the instrument, typically $\sim30~$keV/nucleon, and even solar wind energies (Chotoo et al. 2000).   
More recently, \citet{Ebert2012} studied $73$ CIR-associated suprathermal He intensity enhancements  and found that the peak 
of these sub-MeV He intensities correlate well with the arrival of  the compression region trailing edge. These and 
other recent works (e.g. \citep{Mason.etal08,Bucik.etal09, Ebert.etal12}) suggest 
that sub-MeV energetic particles associated with CIRs at 1 AU may be accelerated locally.

While sub-MeV particles may be accelerated locally, higher energy particles suffer less modulation and   may well be accelerated beyond 1 AU (e.g. $3$ to $5$ AU) and propagate along Parker field back to $1$ AU.  Therefore it is possible that energetic particles observed at 1 AU come  from more than one radial distance. This also implies that cross-field diffusion may be important to understand CIR  observations. Indeed, there is some evidence that cross-field diffusion may be important in  a few CIRs \citep{Dwyer.etal97}.  Because cross field diffusion can drastically reduce the distance a particle travels from several  AUs to reach Earth's orbit, it is therefore more important for low energy particles. 

In this work, we study the transport of energetic particles at CIRs using a numerical model and compare our model with 
an earlier analytical model of \citet{Fisk1980}.  To quantify the modulation effect along a single Parker field,  we assume particles 
are tied to single field lines.  Since there is no cross-field diffusion, particles on  different field lines originate from different 
locations along the reverse shock. 
As a case study, we examine one CIR event that occurred between $2008$ February $08$ and $14$. This event was observed by the twin  Solar TErrestrial RElations Observatory (STEREO) spacecraft and the Advanced Composition Explorer (ACE) spacecraft.  Observations at STEREO-B (hereafter STB) showed an in-situ signature of a reverse shock, which was not seen at ACE and STEREO-A (hereafter STA). This in-situ observation of the shock by STB allows us to estimate the shock location at a later time for   ACE and STB observations upstream of the shock. 

We further assume the shape of the accelerated particle spectrum at the shock does not vary with heliocentric distance. Note that 
this is only a working hypothesis. With this hypothesis, we can decouple the acceleration process from the transport process. This 
hypothesis is bound to introduce some errors in our analysis. However, as we will see, the shock in our event was quite strong at 1 AU,
having a compression ratio $\beta = 2.17$. Inteplanetary shocks often do not have compression ratios larger than, say, $3$. The shock accelerated 
particle spectrum  for a shock with $\beta = 2.17$ is $\sim p^{-5.5}$ while that with a compression ratio $\beta = 3.0$ is \red{$\sim p^{-4.5}$}. 
As we will see, the effect of transport (modulation at low energy) is much larger than the uncertainty introduced in this hypothesis.
Indeed, considering the ratio of the differential intensities  $dJ/dE$(2 MeV)/$dJ/dE$(0.2 MeV) (roughly the energy range for our 
problem),  then the ratio of using $\sim p^{-4.5}$ is larger by a factor of \red{$\sim 3.2$} than that using $\sim p^{-5.5}$. In comparison,
 the effect of modulation on this ratio is more significant. Therefore, our hypothesis, while crude, is somewhat justified. 
Also note that the shock location in our event is not too far from 1 AU. For the two periods we consider, it locates at $1.55$ AU 
and $2.55$ AU, respectively. The plasma parameters at these two shock locations may not differ too much from that at 1 AU and the 
corresponding compression ratios may be similar to $2.17$. \red{In any case}, as we focus on the effects of the transport of energetic 
particles in this work, we here make the simplest assumption about the shape of the accelerated particle spectrum, that is that it 
does not vary with heliocentric distance.

To examine the transport process, we use a Monte-Carlo simulation. In particular, we simulate particle spectra at two 
different times for STB and ACE observations and compare our simulation results with observations. 
In our simulation, we ignore the gyration degree of freedom of the particles and solve the focused transport  equation along a Parker field.  Comparing to \citep{Fisk1980}, who assumed the motion of particles are  diffusive in spatial coordinates and are described by a diffusion coefficient $\kappa$, we retain explicitly  the particle pitch angle in our approach. Consequently, we consider explicitly the magnetic focusing effect and  the pitch angle diffusion (described by $D_{\mu\mu}$) in our formalism.  The \citet{Fisk1980} approach is justified  when the solar wind turbulence is strong. However, when solar wind turbulence is weak, employing $D_{\mu\mu}$ and  treating the focusing effect explicitly is more appropriate.  To summarize,  using a case study where in-situ observation of a CIR-shock is available,  we improve upon the Fisk and Lee model by employing a more sophisticated  numerical simulation that treats the transport of CIR-associated energetic particles by a focused transport equation. 

Our paper is organized as follows. In section \ref{observation} we present the multi-spacecraft observations  of this event; we discuss the simulation model in section \ref{simulation}  and compare the simulation results with observations in section \ref{discussion}.  We conclude in section \ref{conclusion}.

 \section{Event Analysis}\label{observation}

The energetic particle measurements we present here are from the Suprathermal Ion Telescope (SIT) \citep{Mason2008} onboard the STA and STB spacecraft, and the Ultra-low Energy Isotope Spectrometer (ULEIS) \citep{Mason1998} onboard ACE.  The solar wind plasma measurements are obtained from the  PLAsma and SupraThermal Ion Composition investigation (PLASTIC) \citep{Galvin2008}  onboard STA \& B, and the Solar Wind Electron Proton Alpha Monitor (SWEPAM) \citep{McComas1998} onboard ACE.  We also use the magnetic field data from the Magnetometers  on STA \& B \citep{Acuna2008} and ACE  \citep{Smith1998}.

The CIR event we study was first observed at STB starting on $2008$ February $08$.  It has been reported as Event $25$ by \citet{Mason2009}. It has also been studied by \citet{Buvcik.etal11}. 

Figure~\ref{fig:observation} shows the in-situ observations (STB, ACE and STA from left to right). The uppermost panel shows the $0.14$-$2.2$ MeV/n He  time intensity profiles between $2008$ February $07$ and $2008$ February $17$;  the lower panels show, in descending order, the solar wind proton speed, density, temperature and  the total magnetic field (magnitude) in the same time period.  For the STB observations, the He intensity started to increase at $18:00$ UT, $2008$ February $08$ and  reached its maximum at $19:36$ UT, $2008$ February $09$ with a clear peak associated with it.  At the time of the peak, discontinuities in the proton speed, magnetic field and temperature, were observed,  indicating this time as the passage of the CIR reverse shock.  This shock was identified in the STEREO shock list at 
\url{http://www-ssc.igpp.ucla.edu/\~jlan/STEREO/Level3/STEREO\_Level3_Shock.pdf}, maintained by Dr. J. Lan. 
The in-situ shock passage as observed by STB  is shown as the red dashed line in the left panel. The blue and green dashed lines correspond to  two observation periods for which simulations were performed. The shock spectrum is obtained by integrating the differential intensity in the period  $17:54$ UT to $20:09$ UT, February $09$ and is shown by the red curve in Figure~\ref{fig:sourceJ}.  In comparison, the spectrum for a later time period between $13:55$ UT to $15:50$ UT, February $11$ was  shown as the blue curve in Figure~\ref{fig:sourceJ}. The modulation at low energies can be clearly seen for  this period.  

Because the coronal structures from which fast solar wind originated from often evolve slowly,  CIRs are considered as a steady state structure in the co-rotating frame to the first order \citep{Mason2009}. One therefore expects to observe the same reverse shock at ACE after $\sim 1.6$ days.  Indeed, the He intensity observed by ACE began to increase gradually and reached its maximum  at $22:30$ UT, February $10$. However, the proton speed, magnetic field and total pressure showed no  discontinuities, so no reverse shock was observed by ACE.  This is not surprising since time variations of plasma properties such as  density, speed, etc. can cause the location of the shock to vary.  The energy spectrum from ACE observations for the period between $09:59$ UT $\sim$ $12:00$ UT, February $11$ is  shown as the green curve in Figure~\ref{fig:sourceJ}. During this time, ACE is magnetically connected to the shock at a location not far from 1 AU. Consequently, there is little modulation at low energies.

\subsection{Event Geometry}

Figure \ref{fig:stereoloc} shows the relative locations of three spacecraft at $20:00$ UT, 2008 February $09$,  just before the energetic He intensities began to increase on STB. STA, STB and Earth are represented by red (labeled as $A$), blue (labeled as $B$) and green solid circles respectively.  The separation angle between STB and Earth (ACE) is $23.656^\circ$ and the heliocentric distance of STB is 
$1.0$ AU and that of Earth is $0.987$ AU.\footnote{http://stereo-ssc.nascom.nasa.gov/cgi-bin/make\_where\_gif}  Since ACE lies at the L1 Lagrange point between the Sun and the  Earth, its heliocentric distance is $0.977$ AU. 

The configuration of magnetic field lines as well as the CIR reverse shock with respect to STB \& ACE is shown in the cartoon in Figure~\ref{fig:sketch}. The Sun is located at $O$. The black half circle is the trajectory of  STB and the dashed half circle is the trajectory of ACE in the co-rotating frame. Green, blue and red curves  are Parker field lines. The black thick curve extending from $A$ to $D$ represents the CIR reverse  shock surface.  Point $A$ ($B$) is the intersection between the blue (red) magnetic field line and the STB orbit.  Point $H$ ($I$) is the intersection between the green (blue) magnetic field line and the ACE orbit.

When STB is at location $A$, it connects to the local reverse shock, corresponding to observations  at $19:36$ UT, $2008$ February $09$. As STB rotates from $A$ to $B$ in the  co-rotating frame, its connection point at the shock (i.e. the acceleration site) moves further out.  When STB is at location $B$, under the assumption of no cross-field diffusion, it will observe energetic particles that are accelerated by the reverse shock at point $D$. 

\begin{figure}
\noindent
\includegraphics[width=0.8\textwidth]{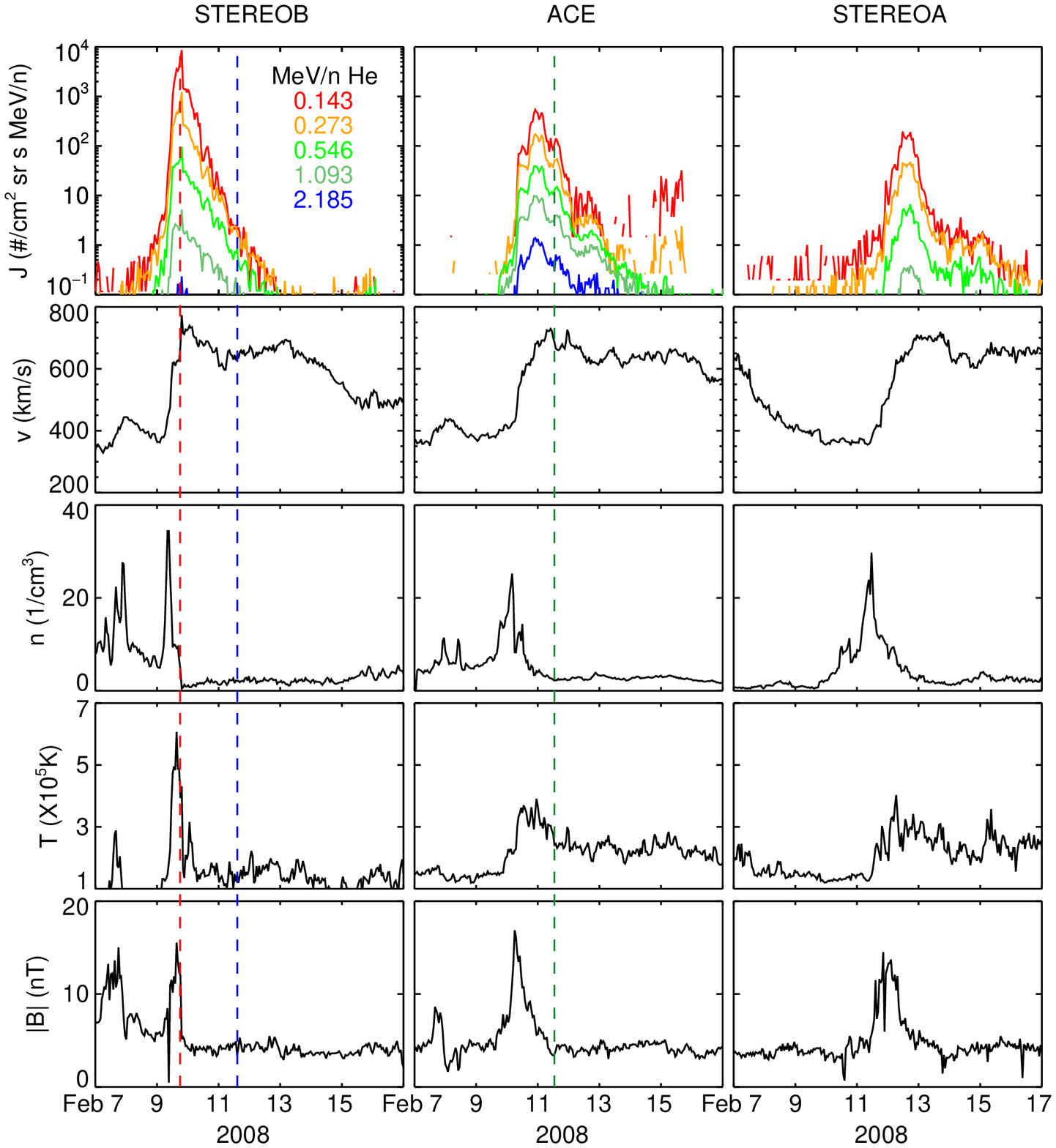}
\caption{CIR observations by STB, ACE and STA (from left to right, respectively) between $2008$ Feburary $07$ 
to $2008$ Feburary $17$. The uppermost panel shows the He intensity time profile, the following panels show 
the solar wind speed, density, temperature, and total magnetic field.
The red dashed line in the left panel marks the shock passage at STB. 
The blue and green dashed lines correspond to  two observation periods for which simulations were performed.
}
\label{fig:observation}
\end{figure}

\begin{figure}
\includegraphics[width=0.5\textwidth]{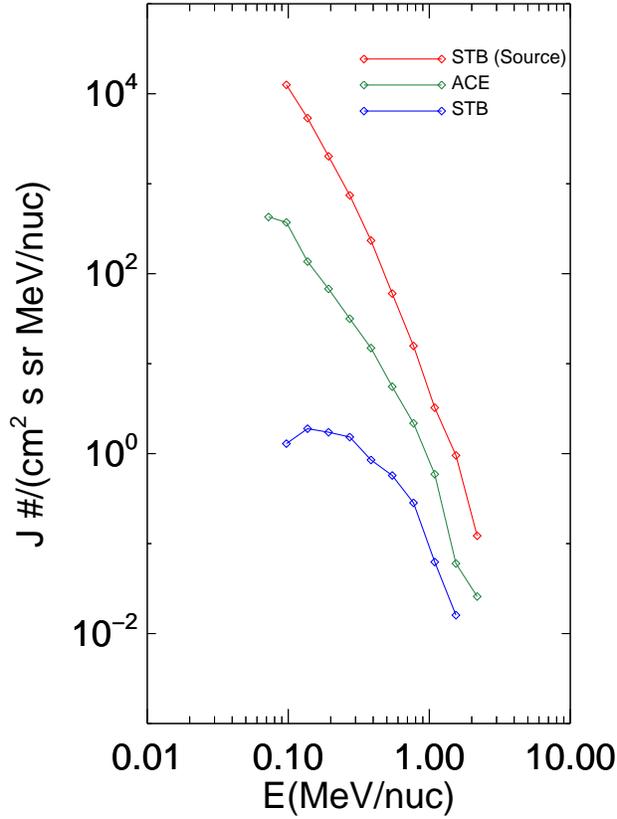}
\caption{Differential intensity spectrum $J(E)$ observed by STB and ACE. Red diamonds show the STB differential intensity 
integrated over $17:54$ UT $\sim 20:09$ UT, Feburary $09$ (corresponding to the red dashed line in Figure~1). 
Blue diamonds show the STB differential intensity integrated 
over $13:55$ UT $\sim 15:50$ UT, Feburary $11$ (corresponding to the blue dashed line in Figure~1). 
Green diamonds show the ACE differential intensity integrated over 
$09:59$ UT $\sim 12:00$ UT, Feburary $11$ (corresponding to the green dashed line in Figure~1).}
\label{fig:sourceJ}
\end{figure}

\begin{figure}
\includegraphics[width=0.5\textwidth]{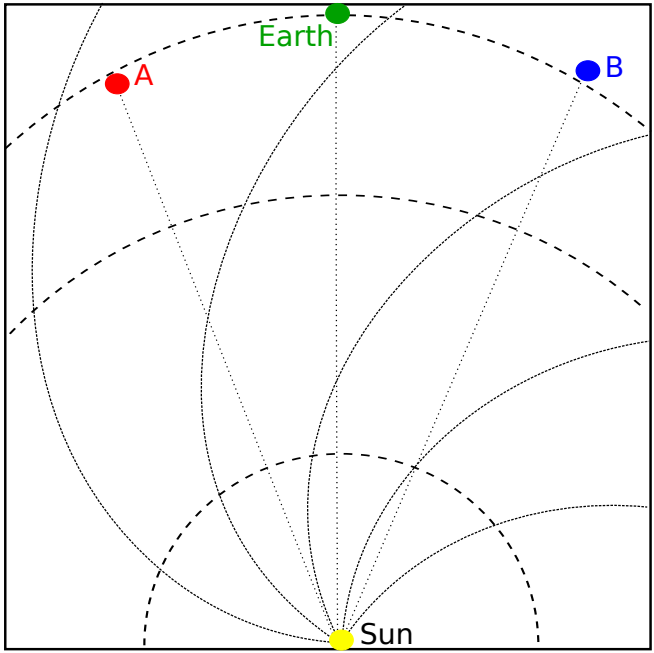}
\caption{Relative locations of STA \& B  and Earth at 20:00 on 2008 Feb 09. STA and STB are indicated by red $A$ and blue $B$. 
Earth and Sun are indicated by green and yellow solid circles. Solid curves are Parker's magnetic field lines. 
Separation angle between $STA$ and Earth is $23.656^\circ$. Separation angle between $STA$ and Earth is 
$21.833^\circ$. Heliocentric distance of STA is $0.966$ AU and that of STB is $\sim 1.0$ AU. }
\label{fig:stereoloc}
\end{figure}

As shown in Figure~\ref{fig:observation}, the intensity of STB He gradually decreases from $19:36$ UT,  February $09$ to  $13:55$ UT, February $11$. This decrease is due to two reasons. First,  as the acceleration sites moves out, it becomes harder and harder for accelerated particles to propagate back  to 1 AU. This is the modulation effect. Second, the density of seed particles, therefore  the intensity of the accelerated particles may also vary with heliocentric distance.   Note, the composition of the seed 
particle is still presently under debate.  In modeling SEP events, where ions and electrons are accelerated at a CME-driven shock  that propagates out from the Sun, it  has been argued and assumed that a fraction of the solar wind  (e.g. $1\%$-$4\%$ as often used in \citet{Li2003, Li2005}) is accelerated at the CME-driven shock.  CIR shocks, however, are often quasi-perpendicular shocks and have higher injection energies than  quasi-parallel CME-driven shocks. 
So it is hard for bulk solar wind to be accelerated. Recently, \citet{Mason2012b}, through a study of the abundance of $^3$He and He$^+$, have suggested that the seed particles for CIR shocks is the  suprathermal ion pool rather than the bulk solar wind.  If the seed particle was the solar wind, then the density of the seed particles will have an $r^{-2}$ dependence.  However, if the seed particle was the suprathermal ions, then the radial dependence of  the seed particle may be more complicated. 
For example, the cooling of solar wind may imply faster decay of the  seed particles than $r^{-2}$. On the other hand, some continuous particle-wave interaction in the solar wind may  suggest a slower decay than $r^{-2}$. Indeed, using Cassini observations,  \citet{Hill.etal09} examined  how the intensities of  $2$-$60$ keV/nuc suprathermal He$^{++}$ vary with $r$. They found that  the intensities of suprathermal He$^{++}$  decrease slower than $r^{-2}$ (see Figure~2 of \citep{Hill.etal09}).

In our event, because STB observed the shock in-situ at 1 AU, if we assume the shock is a steady structure in  the co-rotating frame  and if shock parameters do not change significantly along the shock surface, then with a particle  transport model, we can examine how the intensity of the accelerated particles at the shock varies with radial  distance. 

{From in-situ observation of STB, we find that the average solar wind speed in the fast compression region between 
12:00 $\sim$ 18:00 UT, February $09$ is $v_{comp}=620$ km s$^{-1}$ and in the fast solar wind right after the shock passage is 
$v_{sw} = 760$ km s$^{-1}$ (see Figure~1). Correspondingly, the angle between the upstream magnetic field with the radial 
direction is $\sim 30^\circ$. The reverse shock has a $\theta_{BN}=51^{\circ}$ (see the online list 
\url{http://www-ssc.igpp.ucla.edu/\~jlan/STEREO/Level3/STEREO_Level3\_Shock.pdf}), so the angle between the shock 
normal and the radial direction, $\alpha$, is $20^{\circ}$.} If we approximate the shock normal to be along the radial 
direction, then we can estimate the shock speed $v_{sh}$ at $1$ AU from
\begin{equation} v_{sh}=\frac{\beta  v_{comp}-v_{sw}}{\beta - 1}. \label{Shock_comp}\end{equation}
Here  $\beta=2.17$ is the compression ratio of the reverse shock as obtained from in-situ magnetic field data. 
The calculated shock speed is $500$ km s$^{-1}$ in the s/c frame. 
Note that  $v_{sh}< v_{sw}$ since the reverse shock 
propagates towards the Sun in the fast solar wind frame. Assuming  $v_{sh}$ does not vary  with $r$, we can obtain the radial 
distance $OD$ from, \begin{equation} \frac{CE}{DE} = \frac{v_{sw}}{v_{sw}-v_{sh}}. \label{equ:distance} \end{equation} 
Using Equation \ref{equ:distance}, with the above $v_{sw}$ and {$v_{sh}$}, 
we find  \begin{equation} \frac{CD}{DE} = \frac{CE-DE}{DE}=1.92, \end{equation} and the length of $DE$ 
can be calculated by 
\begin{equation} 
DE = v_{sw} \Delta t, \label{equ:distDE} 
\end{equation} where $\Delta t=1.84$ days 
is the time difference between points $A$ and $B$.  So the distance between $C$ and $D$ is $1.55$ AU. 
Therefore, for STB observation at point $B$, the energetic particles are accelerated at point $D$ on the reverse shock, 
having a heliocentric distance $r_{sh}=2.55$ AU. 
{We remark that the equation~\ref{Shock_comp} is only an approximation and it is only valid when $\alpha$ is 
small (note that $\cos 20^{\circ}=0.94$). 
If the shock normal differs from the radial direction substantially (e.g. $\alpha > 30^{\circ}$, one can not use equation~\ref{Shock_comp}.}

Now consider the ACE observations. Given that the sidereal rotation period of the Sun is $24.47$ days and the angle between 
STB and ACE is $\sim 23.656^\circ$, it takes $1.61$ days for ACE to be connected to the exact same portion of the shock structure as STB, 
implying the enhancement of particle intensity in ACE will be delayed by  $1.61$ days from STB. From the observations shown in Figure 
\ref{fig:observation}, we find a time delay of  the peak intensity between STB and ACE to be $1.14$ days. 
This difference of $d\tau$ between the  observation and by considering solar rotation can be due to a number of  reasons. For example, the ACE and the STB have different latitudes, so ACE and STB did not see the same portion of the shock. 
Finally, the CIR shock may not be completely stationary and its location may vary with time (similar to e.g., the heliospheric termination
shock). The effect of this time variation is illustrated in Figure~\ref{fig:sketch}. In the cartoon, the  dashed thick curve depicts the 
CIR reverse shock when ACE is at point $H$. 

Since ACE did not observe the shock in-situ, it is hard to discern the exact cause of $d\tau$. In the following, we do not consider possible 
oscillations of the CIR shock and  assume that it is given by the thick solid curve as in Figure~\ref{fig:sketch}. 
Following the analysis as in the STB case, we find that  for a later time of ACE observation at $10:00$ UT February $11$, 
the source location at the  shock has a $r_{sh}$ of $1.39$ AU. 

We perform two numerical simulations for two $2$-hour periods corresponding to STB observations  at $13:55$-$15:55$ UT February $11$ and ACE at $10:00$-$12:00$ UT February $11$. The periods are chosen  in which the energetic particle intensities have small variations. We note that STA also observed this event  about 2 days later than ACE.  However, the intensity profile from STA observation showed clear differences from STB and ACE.  
This can be due to the fact that assumptions of a steady-state CIR reverse shock is only applicable within a short duration. 
As shown in \citet{Mason2009} (their Figure 6), energetic particle 
observations of the same CIR event with a few days apart can differ substantially.  
In our event, the separation between STA and STB is $\sim 45^{\circ}$,  $> 3$ days apart, and we do not consider STA observations in this study.

As noted before, we assume the shock is in a steady state in the co-rotating frame and that  the shock strength does not vary along the shock surface. Assuming the shock parameters do not vary as a  function of heliocentric distance allows us to use the 1 AU in-situ shock spectrum from STB as a  reference for energetic particle spectra at different times. Therefore, as noted before, we can decouple the  transport process from the acceleration process for CIR-associated energetic particles.  

Specifically, we will use the in-situ STB 1 AU shock observation of the energetic particle  differential current intensity $J(E)$ and scale it by a factor of $\eta$ (defined below) at two different locations at the CIR reverse shock, to calculate the resulting differential intensities as  observed at 1 AU by STB and ACE. By comparing simulations with observations we can obtain how $\eta$,  which is a measure of the seed particle intensity, varies with $r$.  The implication of this radial dependence of $\eta$ is discussed in section~\ref{discussion}. 

 \section{Model Description}\label{simulation}

We use a Monte-Carlo code to study the transport of charged particles.  The transport of energetic particles can be described by the focused transport equation \citep[e.g.][]{Skilling1971,Isenberg1997}: \begin{equation} \begin{split} \frac{\partial f}{\partial t} + (u_i + v\mu b_i)\frac{\partial f}{\partial x_i}  + [\frac{1-3\mu^2}{2}b_ib_j\frac{\partial u_i}{\partial x_j} - \frac{1-\mu^2}{2}\frac{\partial u_j}{\partial x_j}  - \frac{\mu b_i}{v}(\frac{\partial \mu}{\partial t}+\frac{\partial u_i}{\partial x_j})]v\frac{\partial f}{\partial v} \\ + \frac{1-\mu^2}{2}[v\frac{\partial b_i}{\partial x_i } +  \mu\frac{\partial u_i}{\partial x_i}-3\mu b_i b_j\frac{\partial u_j}{\partial x_i}- \frac{2b_i}{v}(\frac{\partial u}{\partial{t}}+u_j\frac{\partial u_i}{\partial x_j})] \frac{\partial f}{\partial \mu} = \frac{\partial}{\partial \mu}D_{\mu\mu}\frac{\partial f}{\partial \mu} + S - L. \end{split} \label{equ:focused} \end{equation} 
In the above, the distribution function $f$ is assumed to be gyrotropic; the $b_i$ and $u_i$ are the components of the unit magnetic field and solar wind speed. $\mu$ is the particle's pitch  angle cosine. $S$ and $L$ are the source and loss terms. The focused transport equation can be derived from the  guiding center theory \citep{LeRoux2009}. The particle's motion therefore can be   regarded as the motion of the guiding center along the background magnetic field line plus a diffusion in pitch  angle, governed by $D_{\mu\mu}$.

In the solar wind, the background magnetic field is given by 
\begin{equation} 
\begin{split} 
B_r = &B_0(\frac{R_0}{r})^2,\\ 
B_\theta= &0, \mbox{and}\\ 
B_\phi = &B_r\frac{\Omega r \sin \theta}{u}  \:(r >> R_0) \end{split} \label{equ:spiralB} \end{equation} 
where $\theta = 90^\circ$ corresponds to the magnetic field in the ecliptic plane.  In this work, we use $B_0 = 1.53$ Gauss at 
$R_0 = 1$ $R_s$ (solar radius).

Energetic particles are followed in two different frames \citep[see][]{Ruffolo1995,Agueda2005,Kocharov2003}.  
One is the instantaneous co-rotating frame,  the other is the instantaneous solar wind frame. The instantaneous co-rotating frame is a frame which co-rotates  with the Sun. The instantaneous solar wind frame is a local inertial frame that co-moves with the solar wind at the  particle's location.  In the instantaneous co-rotating frame, the solar wind velocity is $v_{sh}^{co}=v_{sw}/\cos\psi$,  where $\psi$ is the angle between the magnetic field direction and the radial direction.  The solar wind velocity is then parallel to the local magnetic field, yielding a zero induced electric field.  Therefore, the particle's energy is conserved in this frame. 
Furthermore, due to the focusing effect, the  particle's pitch angle will change because of the conservation of the particle's magnetic moment $p_\perp^2/B$.  At each time step, the particle's location and pitch angle are updated in this instantaneous co-rotating frame.  We then transform the particle's momentum to the instantaneous solar wind frame. This is the frame  where the effect of solar wind Magnetohydrodynamic (MHD) turbulence is considered since MHD waves and turbulence  are generated locally. Pitch angle diffusion is considered in this frame. After considering pitch angle diffusion, we transform back to the instantaneous co-rotating frame.  Finally, we need to transform the particle's energy and momentum between two instantaneous co-rotating frames  at two different times $t$ and $t$ + $\delta t$. As shown in Appendix A, to the order of $(v/c)^2$, the particle's energy   is conserved under the Lorentz transformation between these two frames.

 In modeling the pitch angle diffusion, we follow \citet{Zhao2014}.  For each time step $\delta t$, the particle's pitch angle $\mu$ will change  by a small amount $\delta \mu$ given by \begin{equation} \delta \mu = sign(\epsilon ' - \frac{1}{2})erf^{-1}(\epsilon) \sqrt{4D_{\mu\mu}\delta t} +\frac{\partial D_{\mu\mu}}{\partial t}\delta t, \end{equation} here $\epsilon'$ and $\epsilon$ are random numbers uniformly distributed in $[0, 1]$. $erf^{-1}$  is the inverse of error function.  In Quasi-linear theory (QLT) \citep{Jokipii1971, Lee1974,Wentzel1974}, $D_{\mu\mu}$ is given by \begin{equation} D_{\mu\mu}=\frac{1}{2}\frac{<(\delta \mu)^2>}{\delta t} = \frac{\pi}{4}(1-\mu^2)\Omega_o\frac{kP(k)}{B_o^2}, \end{equation} where $P(k)$ (given in Equation \ref{equ:pk}) is the turbulence power spectrum in solar wind.  $\Omega_o = eB/(\gamma m)$ is particle's gyrofrequency. $k = \Omega_o(v|\mu|)^{-1}$ is particle's resonant wave number. 
\begin{equation} P(k) = A_q \lambda_c(\delta B)^2(1+(k\lambda_c)^q)^{-1} \label{equ:pk} 
\end{equation} 
$\lambda_c$ is the correlation length and $q$ is the power law index of the turbulence spectrum. $A_{\beta}$ is  determined by the normalization condition 
\begin{equation} \int_{k_s}^{k_L} P(k)dk = (\delta B)^2. \end{equation} 
The correlation length $\lambda_c$, smallest ($k_S$) and largest ($k_L$) wave number in $P(k)$ are set to be the typical 
values at $1$ AU \citep[see][]{Zhao2014}.

 The radial dependence of $B$ is well defined for a Parker field. For $\delta B$, however, the radial dependence is still presently unknown. Some earlier studies by \citet{Bruno2005} suggested that a WKB approximation, i.e., $\delta B^2 \sim r^{-3}$ may be reasonable but slightly underestimated. Following \citet{Mason2012a}, we  use $\delta B^2 \sim r^{-3.5}$ in this work, so \begin{equation} \delta B^2 (r) = \delta B^2(1 {\rm AU})  (\frac{r}{1 {\rm AU}})^{-3.5}  \end{equation}

We assume the shape of the source energetic particle spectrum $f(E)\sim J(E)/v$ does not vary with $r$, and is  given by a broken power law form as shown in Figure~\ref{fig:source}.  The form of $f(E)$ is chosen to fit the observed differential intensity $J(E)$ at STB  immediately downstream of the shock between $17:54\sim20:09$ UT, February $09$.   In Figure~\ref{fig:source}, the low-energy portion of $f(E)$ (shown in blue) has  $f(E)\sim E^{-3.57}$ between $0.094 < E/({\rm MeV/nucleon}) <0.546$, and the high-energy portion  (shown in red) has $f(E)\sim E^{-4.89}$ between $0.546 < E/({\rm MeV/nucleon}) <2.185$.

For our simulation, we inject a total of $N_0 = 60,000$ protons at location of $r_{sh}$ (which is also the outer  boundary) 
with an initial pitch angle cosine $\mu$ uniformly distributed between $-1$ and $0$.  
{We do not consider the acceleration process in this work. So these particles are injected with a given spectrum. 
We assume this spectrum, up to a factor, is the same as that from the in-situ observation of the shock at STB. In this way,
 we attempt to decouple the acceleration process from the transport process. Note that close to the shock, the acceleration 
process is governed by the diffusion cofficient, which can be substantially smaller than that in the interplanetary medium. So
 the acceleration process is quite different from the transport process. We release all particles at the same time and 
follow all of them for a period of $10$ days and obtain the time-integrated spectrum. 
We then follow the transport of these particles from the shock to 1 AU. }
Particles leave the simulation domain when they reach either the inner boundary ($r=0.01$ AU) or the  outer boundary ($r=r_{sh}$). 
The particles' momenta and pitch angles are recorded when they pass $r=1$ AU. 

\section{Results and Discussion}\label{discussion}

The simulated proton differential intensities are shown in Figure~\ref{fig:result} as the black curves with  ``diamond'' symbols for two cases. The left panel is for case I, corresponding to the STB observation  between $13:55$ UT $\sim 15:50$ UT, February $11$.  In this case, energetic protons are injected at a heliocentric distance $r=2.55$ AU and  the particle differential current is obtained at $1.0$ AU. The observed differential current  intensity is shown as the blue curve with the ``plus'' sign.  The right panel is for case II. In this case, energetic protons are injected at a heliocentric distance  $r=1.39$ AU and the particle differential intensity is obtained at $0.98$ AU. This case corresponds to the ACE  observation between $10:00$ UT $\sim12:00$ UT, February $11$.  The observed differential intensity is shown as the green curve with the ``plus'' sign.  In both cases, we use a two-hour observation window so that enough statistics can be obtained.  We assume the shock location does not change during this $2$-hour interval. We do not consider a  longer period ($>$ 2 hours) because the acceleration site to which the spacecraft is connected to can change  rapidly along the shock surface.

To fit the observed differential intensities at 1 AU in Figure~\ref{fig:result},  a key parameter is $(\delta B/B)^2$ at 1 AU. We calculate $(\delta B/B)^2$ using the STB/MAG 1-minute data and the ACE/MAG 4-minute data. In the case of STB observation, the average magnetic field was obtained for a period of 20 hours before the 2-hour observation window  and we find $(\delta B/B)^2 = 0.010$. For the ACE observation, the background magnetic field before  the 2-hour observation window showed a clear decreasing trend, so we have chosen a 20-hour period after  the 2-hour observation window to calculate the average magnetic field and  $(\delta B/B)^2$.  We find $ (\delta B/B)^2 = 0.019$ in this case. The magnetic field data and the periods for calculating  the average background magnetic field and $(\delta B/B)^2$ for both the ACE and STB observations are  shown in figure~\ref{fig:avgB}. In the figure, the two left panels are for STB and the two right panels are for ACE.  In both cases, the upper panels show the total $B$ for an extended period covering the whole event and the  lower panels are zoom-in plots of the total magnetic field and $(\delta B/B)^2$ for a 20-hour  period before (for the case of STB) or after (for the case of ACE) the 2-hour observation window of  energetic particles. 

Now consider the fitting of STB observation in Figure~\ref{fig:result}.  We vary the ratio of  $(\delta B/B)^2$ at $1$ AU from the observed in-situ value of $0.01$ to fit the shape of the observed differential  intensity at 1 AU.  Note that the observed $(\delta B/B)^2 =0.01$ is an ensemble average of many radially propagating  plasma parcels that pass through the spacecraft during a $20$ hour window. These plasma parcels do not consist of a  Parker field line on which energetic particles propagate from $2.55$ AU to $1$ AU. So it is only a proxy of the  turbulence level along the Parker field line particles propagate on.  Nevertheless, the best fit yields a $(\delta B/B)^2 =0.012$ at $1$ AU, very close to the in-situ observation. The simulated and observed differential current intensities for STB observation agree nicely, as can be seen  from the left panel of Figure~\ref{fig:result}.  Next consider the ACE observation. Again, by varying the ratio of  $(\delta B/B)^2$ at $1$ AU, the best fit yields $(\delta B/B)^2=0.02$,  close to the in-situ observation of  $(\delta B/B)^2 =0.019$ at ACE.  This is shown in the right panel of Figure~\ref{fig:result}. 

After fitting the spectral shape, we next obtain the normalization factor $\eta$ of the injected particles from the fitting. 
The parameter $\eta$ is defined through, 
\begin{equation} \eta= \frac{\int^{E_{\rm max}}_{E_{\rm min}}  dE J_{sh} } { \int^{E_{\rm max}}_{E_{\rm min}} dE J_{1{\rm AU}} } \end{equation} where $J_{sh}$ is the differential intensity at the shock ($r=1.39$ AU for the ACE observation and $r=2.55$ AU  for the STB observation) and $J_{1{\rm AU}}$ is the differential current when the shock is  observed in-situ at 1 AU by STB between $17:54$ $\sim$ $20:09$ UT, February $09$. The parameter $\eta$ reflects  how the number of seed particles, i.e., the particles that participated in the shock acceleration process, depends on  $r$. In the simulation, we inject particles to a flux tube whose cross section depends on the shock location.  
{Because the footpoint of the flux tube on the shock surface varies with heliocentric distance, we
introduce an additional parameter $\alpha$ through,}  
\begin{equation} \alpha = \eta \frac{A_{sh}}{A_{1AU}}   \label{eq:alpha} \end{equation} where $A_{sh}$ is the cross section of the flux 
tube that intersects with the shock at $r$, and $A_{\rm 1 AU}$ corresponds  to the in-situ STB observation of the shock at 1 AU. 
The factor $A_{sh}/A_{\rm 1 AU}$ reflects how the flux tube  expands as a function of $r$. Knowing the solar wind speed and the shock speed,  the ratio   $A_{sh}/A_{1AU}$ can be readily calculated as shown in Appendix B. For our cases, $A_{1.39 AU}/A_{1AU} = 1.65$, and  $A_{2.55 AU}/A_{1AU} = 4.72$. From the fitting we obtain $\alpha = 6$ for case I and $\alpha = 1$ for case II. Therefore we obtain $\eta_{\rm case I} =1.27$ and $\eta_{\rm case II} =0.61$. Our Simulation results are summarized in Table~\ref{tab:MonteCarloFitting}.

These values of $\eta$'s are very important results.  Consider first the ACE observation with $r_{sh}=1.39 {\rm AU}$. In this case we find that to fit the observation,  the number of accelerated particles at $r_{sh}$ is $0.6$ times that at $1$ AU. This decrease yields a radial dependence  of $r^{-1.55}$. This is shallower than the radial dependence of solar wind density $r^{-2}$, and  is consistent with  \citet{Hill.etal09} where the intensities of suprathermal He$^{++}$ was found to  decrease slower than $r^{-2}$. Therefore the ACE observation supports the notation that the seed particles for  CIR-associated energetic particles are most likely suprathermal ions than the bulk solar wind \citep{Mason2012b}.

Comparing to the ACE observation, the STB observation with $r_{sh}=2.55 {\rm AU}$  shows that $J_{sh}(r= 2.55 {\rm AU})/J_{sh}(r=1 {\rm AU}) = 1.27$. So the number density of the accelerated  particles at $2.55$ AU has to be larger than that at 1 AU. This implies that the intensity of the seed particles  increases with $r$, instead of decreasing with $r$. It contradicts with the results in \citep{Hill.etal09}.  However, we note that the data points in \citep{Hill.etal09} are $0.5$-$1$ AU apart, and most are far beyond  the Earth orbit, so they may not provide much constraint about the relatively small shock distances studied here.  A larger seed population further out than 1 AU may seem counter-intuitive. However,  earlier works \citep{McDonald1976, VanHollebeke1978} have shown that CIR intensities  increase beyond 1 AU and peak at several AU. Note that in these earlier studies, the enhancement of CIR  intensity is largely attributed to the fact that CIR shocks tend to form beyond 1 AU  (e.g. \citet{Desai.etal99}). In our case, however, the CIR shock was seen to form in-situ at 1 AU by STB.  Note that we have assumed the shock parameters do not vary with the heliocentric distance of the shock. It is  possible that the shock strengthens beyond 1 AU in that the injection energy decreases with r, so that more  particles can participate in the shock acceleration process. This, of course, still implies that the seed population  (i.e. particles participating the shock acceleration process) increases with $r$.

This increase of seed particle intensity at $2.55$ AU is required because the modulation effect for low energy  particles is significant.  This can be seen from the following arguments.  Under the assumption that particles do not diffuse across the field lines, the path  length from $2.55$ ($1.39$) AU to $1$ AU is $2.26$ ($0.52$) AU (assuming a solar wind speed of $760$ $km~s^{-1}$).  If there is no scattering ($\delta B/B_0 \rightarrow 0$), then as a charged particle moving towards  the Sun along the Parker field, the focusing effect (conservation of particle's magnetic moment) will tend to reverse  the particle's momentum direction. With turbulence included, the pitch angle will undergo both a focusing and a scattering process. To compete with the focusing effect, one may think that a larger $\delta B/B_0$ will help.  However, if $\delta B/B_0$ is too large so that pitch angle scattering dominates focusing, then the motion of  particles along the field can be regarded as a diffusion and the time to arrive 1 AU from the shock will be very long.   For example, if the mean free path of a $0.5$ MeV proton is $\sim 0.5$ AU (see e.g. equation (27) of \citet{Li2003}),  it takes $\sim 10$ mean free paths for an $0.5$ MeV proton to arrive 1 AU from $2.55$ AU, translating to a time period  of $21$ hours. Since the adiabatic cooling rate is $\nabla \cdot v_{sw}$, the longer the propagation time,  the more deceleration a particle will experience, leading to a stronger modulation. If the mean free path (mfp) $\lambda$ is smaller, since the travel time scales as $\lambda^{-1}$,  the adiabatic cooling will lead to a stronger modulation.  Finally,  for particles of smaller energies, using equation (27) of \citet{Li2003} (with $\alpha=1/3$), one finds that  the travel time scales roughly as $E^{-2/3}$, so the modulation will be 
even stronger.

 The above discussion illustrates why a significant number of seed particles has to be present when the shock is further out.  Note that one implicit assumption in our scenario is that  particles are tied to a single magnetic field line.  So we have ignored $\kappa_{\perp}$.  The picture can be changed, and perhaps substantially,  if we allow particles to diffuse across field lines. This can be seen easily from Figure~\ref{fig:sketch}.  Within a small longitudinal range we see that the shock goes from $1$ AU (point A) to $2.55$ AU (point D).  While it is difficult for  low energy particles accelerated at point D ($2.55$ AU) to propagate along the Parker spiral to point B ($1$ AU), if the low energy particles could cross-field diffuse from field lines that connect to the shock at a closer distance  than $2.55$ AU (e. g. field lines to the left of $BD$) then their propagation distance would be much smaller . Consequently, if cross-field diffusion plays  an important role, then we do not need to require as large a seed population further out. We note that if particles could cross-diffuse in to, e.g. the field line $BD$ as shown in Figure~\ref{fig:sketch},  particles on the field line of $BD$ could also cross-diffuse out. However, since low energy particles accelerated at  point $D$ will have a hard time to propagate to point $B$, so whether they stay in the same field line or  diffuse out to other field lines does not matter much to observations at point $B$.  

 It is instructive to fit the differential intensities for both case I and case II using the Fisk \& Lee model.  Following \citet{Fisk1980}, one can show that the upstream differential intensity $J(E)$ is,  
\begin{equation} J  = J_0(\frac{R}{R_s})^{2 /(\beta-1) + V/(\kappa_0 v)} v^{n} e^{-v/v_0}  \label{eq:FiskLeeDownstream}   
\end{equation} 
where 
\begin{equation} 
\beta = (\frac{V^2_{up} +\Omega^2R_s^2}{V^2_{dn} +\Omega^2R_s^2})\frac{B_{dn}}{B_{up}} \label{eq:FiskLeeCompRatio} 
\end{equation} 
In the above $\beta$ is the compression ratio, $n = -(\beta+2)/(\beta-1)$;  $v_0=\frac{V(\beta-1)^2}{6 \kappa_0 \beta}$; 
$V$ the solar wind speed; $\Omega$ the angular rotation speed of the Sun; subscripts ``up'' and ``dn'' refer to quantities 
upstream and downstream of the CIR reverse shock; and $J_0$ is a fitting constant.  At the CIR shock,  $J = J_0 v^{n}e^{-v/v_0}$.

There are four free parameters in equation~(\ref{eq:FiskLeeDownstream}).   These are $R_s$, $J_0$, $\beta$ and $v_0$. 
In fitting the observed CIR-associated energetic particle spectrum using the functional form of 
 equation~(\ref{eq:FiskLeeDownstream}), one issue is that for rather different choices of the parameter set 
($J_0$, $\beta$ and $v_0$),  one can obtain very comparable fitting results. If, however, we assume that the solar wind speed 
and the shock compression ratio do not vary as a function of heliocentric distance, then $\beta$ is fixed by the STB in-situ 
observation. Furthermore, $v_0$ is also fixed by the STB in-situ observation. Therefore, the only two free parameters are 
$J_0$ and $R_s$.

Figure~\ref{fig:FiskLeeFitting} shows the fitting results for the differential intensities for,  
1) STB in-situ observation of the reverse shock 
during $17:54$ UT $\sim 20:09$ UT, Feburary $09$, 2008 2)  ACE observation during  $02:30$ UT $\sim 04:30$ UT, 
Feburary $11$, 2008  and 3) STB observation during  $0:00$ UT $\sim 02:00$ UT, Feburary $11$. To obtain the 
best fitting, hese are slightly different time periods from our simulation.

Fitting the differential intensity at the reverse shock as observed by STB using a compression ratio $\beta = 2.13$ from the in-situ measurement,  we find $J_0$ (in unit of cm$^{-2}$ s$^{-2}$ sr$^{-1}$ (MeV/n)$^{-1}$)   is $1227$ 
and $v_0 = 0.172$ (MeV/n)$^{1/2}$  (corresponding to $\kappa_0 = 0.03$). Using these two parameters we then 
find  $J_0$(ACE)$=4167$ and $R_s = 1.73$ AU; $J_0$(STB)$=5474$ and $R_s = 3.63$ AU. 
The fitting results of $J_0$ and $R_s$ are summarized in Table~\ref{tab:FiskLeeFitting}. These shock locations are to be 
compared with our simulations where $R_s=1.39$ AU for the ACE observation and  $R_s=2.55$ AU for the STB observation. 
The arbitrariness of the fitting can be seen from the following: for the ACE fitting, for example, we can obtain
an almost equally well fitting  with $J_0$(ACE)$=2070$  and $R_s = 1.5$ AU.

In any events, we see that the Fisk \& Lee model yields somewhat larger shock locations for both periods.
Now considering the differential intensities: the Fisk \& Lee model predicts a differential intensity at $R_s = 1.73$ AU to  
be $\frac{4167}{1227}=3.4$ times that when $R_s=1$ AU (or a differential intensity at $R_s = 1.5$ AU to  
be $\frac{2070}{1227}=1.69$ times that when $R_s=1$ AU); and a differential intensity at $R_s = 3.63$ AU to be  
$\frac{5474}{5160}=4.7$ times that when $R_s=1$ AU. Therefore the Fisk and Lee model also suggests that the differential 
intensity at the shock increases with heliocentric distance of the shock. 

\section{Conclusion}\label{conclusion} 
We have examined energetic particle transport in one CIR event which occurred in $2008$ February.  We choose this event because the CIR was observed by both STB and ACE and the CIR-associated  energetic particle intensity profiles from these two spacecraft reasonably resemble each other  (with a time shift of $\sim 1.4$ days). Furthermore, STB observed the reverse shock in-situ,  suggesting that the shock was formed near $1$ AU. Under the assumption of no cross-field diffusion, we develop a Monte-Carlo test particle model to investigate the  transport of energetic particles. The model solves the focused transport equation numerically.  Comparing to previous analytical work by \citet{Fisk1980}, our model considers explicitly the particle pitch angle evolution.   Both the magnetic focusing effect and the pitch angle diffusion process are included. For the cases where the solar wind  MHD turbulence is not strong, our approach is more appropriate than that of  \citep{Fisk1980}.

Assuming the reverse shock can be approximated by a stationary structure in the co-rotating frame, and that the  shape of the accelerated particle spectrum at the shock does not vary with heliocentric distance, we calculated the different intensity at $1$ AU for two periods corresponding to an ACE observation and  a STB observation.  
{By assuming the accelerated particle spectrum at the shock 
is given by that observed in-situ at STB and \red{does} not vary with heliocentric distance, we avoid considering particle acceleration at 
the shock explicitly. Presumably, the acceleration at the shock \red{does} vary with the heliocentric distance. Under the diffusive 
shock acceleration framework, the accelerated particle spectrum depends on various shock parameters including the shock geometry 
and the shock compression ratio. Both can depend on the heliocentric distance. However, since we have only 1 AU observations, no 
constraints on these parameters can be obtained. We note this as a limitation of the present work. }

By assuming a turbulence similar to that given by the WKB approximation, reasonable agreements between the simulation and the  observations are obtained for both observations. The best fit of the STB observation yields $(\delta B/B_0)^2 = 0.012$, similar to  the in-situ value of $0.01$. The best fit of the ACE observation yields $(\delta B/B_0)^2 = 0.02$, also  similar to the in-situ value of $0.019$.  The ACE fitting suggests that the seed particle density at $r_{sh}=1.39$ AU  is $0.61$ times that when $r_{sh} = 1$ AU, consistent with the radial dependence of suprathermal He$^{++}$ obtained in  \citet{Hill.etal09}. However, the STB fitting suggests that the seed particle density  at $r_{sh}=2.55$ AU has to be $1.27$ times larger than that when $r_{sh} = 1$ AU. This contradicts to \citet{Hill.etal09}.  This requirement of a large seed particle density at $r_{sh}=2.55$ AU is due to the fact that the modulation, especially  for low energy particles, is strong. 

This contradiction may be resolved by including cross-field diffusion. Including cross-field diffusion can effectively  negate the modulation effect since with cross-field diffusion included, low energy particles that accelerated at  a shock location closer to $1$ AU can diffuse cross-field and contribute to the observed intensity at a location  that magnetically connects to the shock at a large distance (e. g. $\sim 2.55$ AU).

In summary, we have developed a numerical Monte-Carlo code to examine energetic particle transport at CIR shocks. We ignored  particle cross-field diffusion and consider particles propagating along a single field line.   We examined one CIR event which occurred in $2008$ February using our code. We simulated energetic particle spectra for  two two-hour windows at both ACE and STB.  Very good agreement between the simulation and observation can be obtained if we assume the seed particles at $2.55$ AU is $\sim 1.27$ times that at 1 AU and if the seed particles at $1.39$ AU is $\sim 0.61$  times that at 1 AU. 
However, we note that this conclusion may be changed if cross field diffusion is included.

\begin{acknowledgments} 
{We thank the many individuals at the University of Maryland, Johns Hopkins Applied Physics Laboratory, and
Goddard Space Flight Center responsible for the construction of the ACE/ULEIS and STEREO/IMPACT/SIT instruments.
We acknowledge the use of data from the STEREO and ACE missions. 
The STEREO/SIT data was obtained from \url{http://www.srl.caltech.edu/STEREO/Level1/SIT\_public.html},
while the STEREO/PLASTIC and /MAG data was obtained from 
\url{http://aten.igpp.ucla.edu/forms/stereo/level2\_plasma\_and\_magnetic\_field.html}. 
The ACE/ULEIS, /SWEPAM, and /MAG data was obtained from the ACE Science Center 
(\url{http://www.srl.caltech.edu/ACE/ASC/level2/}).  }
GL and LZ were supported in part by NSF grant AGS1135432, NASA grants NNX13AE07G, and NNX14AC08G at UAH; 
RWE, MID, and MAD at SWRI were supported in part by NASA grant NNX13AE07G,
NSF grant AGS-1460118 and an internal research grant from Southwest Research Institute. The work at APL was 
supported by NSF grant AGS-0962653, contract SA4889-26309 and NASA grant NNX13AR20G from the University of California 
Berkeley. The work at ShangDong University was supported by  NNSFC grants 41274175 and 41331068. \end{acknowledgments}

\appendix \section{Instantaneous Co-rotating Frame Transformation} The adiabatic deceleration effect in an expanding solar wind is described by \citep{Parker1965, Jokipii1970} \begin{equation} <\dot{p}> = -\frac{p}{3}\nabla\cdot v_{sw} \end{equation} where $v_{sw}$ is the solar wind speed and $p$ is particle's momentum in solar wind frame. This expression of average deceleration rate is valid under the assumption that particle's pitch angle diffusion is nearly isotropic \citep{Ruffolo1995}. In \citet{Ruffolo1995} the analytical expressions of the particle's momentum (pitch angle) deceleration rate for an individual particle is given by \begin{equation} \begin{split} \dot{p} &= -pv_{sw}[\frac{sec\psi}{2L(z)}(1-\mu^2)+\cos\psi\frac{d}{dr}sec\psi\mu^2]\\ \dot{\mu}&=\frac{v}{2L(z)}[1+\mu\frac{v_{sw}}{v}sec\psi-\mu\frac{v_{sw}v}{c^2}sec\psi](1-\mu^2)-v_{sw}(cos\psi\frac{d}{dr}sec\psi)\mu(1-\mu^2) \end{split} \label{equ:adiabatic} \end{equation} where $\mu$ denotes the particle's pitch angle in solar wind frame. $\psi$ is the angle between the radial direction and the magnetic field's tangent direction. 
$1/L(z) = -B(z)/(\partial B/\partial z)$ is the reciprocal of the scale length of the interplanetary magnetic field. The first equation in Equation \ref{equ:adiabatic} describes the adiabatic cooling effect and the second equation describes pitch angle focusing effect.

In this paper, particle's adiabatic deceleration and focusing effect are not modeled by Equation \ref{equ:adiabatic}, instead these effects are treated implicitly by a frame transformation approach. Figure \ref{fig:frametrans} shows the local coordinates of two adjacent instantaneous co-rotating frames. In panel ($a$), the black curve is the spiral magnetic field line and $A$, $B$ are two adjacent locations with a time interval $dt$. The radial distance of $A$ is $R_1$ and of $B$ is $R_2$. $\bm{r_1}$ ($\bm{r_2}$) is the unit vector in the direction of $R_1$ ($R_2$).  $\bm{\theta_1}$ ($\bm{\theta_2}$), which is also the direction of instantaneous co-rotating frames at location $A$ ($B$), is unit vector in the direction perpendicular to $\bm{r_1}$ ($\bm{r_2}$). The four vectors' relative directions are shown in panel ($b$). $\Delta \psi$ is the angle between $\bm{r_1}$ and $\bm{r_2}$. 

 Suppose the Sun rotates with an angular velocity $\Omega$, we get the velocities of the two instantaneous co-rotating frames 
\begin{equation} 
\begin{split} \bm{V_1} =&\Omega  R_1 \:\bm{\theta_1},\\ 
\bm{V_2} =&\Omega  R_2 \:\bm{\theta_2}, \end{split} \label{equ:vel} 
\end{equation} 
and the relative velocity between two co-rotating frames is 
\begin{equation} 
\bm{V_1} - \bm{V_2} = \Omega R_1 \:\bm{\theta_1} - \Omega R_2\:\bm{\theta_2}.
 \end{equation} 
Take the dot product of ($\bm{V_1}-\bm{V_2}$) and $\bm{AB}$ and substitute Equation \ref{equ:vel}, we get 
\begin{equation} \begin{split} &(\bm{V_1} - \bm{V_2})\bm{\cdot}(R_1\:\bm{\theta_1} - R_2\:\bm{\theta_2}) \\ 
=&\Omega  R_1^2 \:\bm{\theta_1}\bm{\cdot}\bm{r_1} + \Omega R_2^2 \:\bm{\theta_2}\bm{\cdot}\bm{r_2} -\Omega  
R_1 R_2 \:\bm{\theta_1}\bm{\cdot}\bm{r_2} - \Omega R_2 R_1 \:\bm{\theta_2}\bm{\cdot}\bm{r_1}. \end{split} \label{equ:relavel} 
\end{equation} 
Clearly from panel ($b$) in Figure \ref{fig:frametrans}, $\bm{r_1}$ is perpendicular to $\bm{\theta_1}$ and $\bm{r_2}$ 
is perpendicular to $\bm{\theta_2}$, then \begin{equation} \begin{split} \bm{r_1}\bm{\cdot}\bm{\theta_1} &= 0,\\ 
\bm{r_2}\bm{\cdot}\bm{\theta_2} &= 0,\\ \bm{r_1}\bm{\cdot}\bm{\theta_2} &= \sin(\Delta\psi),\\ 
\bm{r_2}\bm{\cdot}\bm{\theta_1} &= -\sin(\Delta\psi).\\ \end{split} \label{equ:dotpro} 
\end{equation} 
Combining Equation \ref{equ:relavel} and \ref{equ:dotpro}, the dot product of ($\bm{V_1}-\bm{V_2}$) and $\bm{AB}$ is zero, 
which means if the time step is small, the relative velocity between two co-rotating frames is perpendicular to the Parker's 
spiral (particle's trajectory). 
{One can then obtain the particle's energy and the momentum component parallel and perpendicular 
to the Parker's spiral from the Lorentz transformation.} 
%
%

\section{Determination of Shock Cross Section}\label{app:cross} 
\begin{figure}
\includegraphics[scale=0.5]{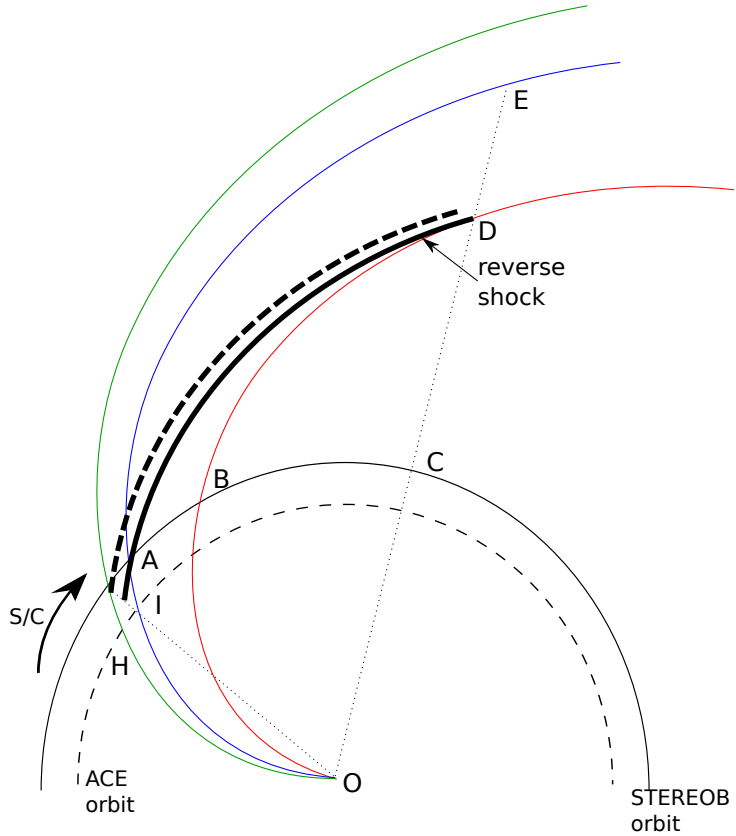}
\caption{Configuration of CIR with respect to STB \& ACE. The Sun is located at $O$, black half circle is trajectory of STB 
in co-rotating frame and dashed half circle is trajectory of ACE. Green, blue and red curves are Parker's spiral magnetic 
field lines. $A$($B$) is the intersect point of blue (red) curve and STB's orbit. $H$ ($I$) is the intersect of green (blue) 
curve and ACE orbit. Black thick curve (dashed thick curve) extending from $A$ to $D$ is reverse shock surface.}
\label{fig:sketch}
\end{figure}

\begin{figure}
\includegraphics[width=0.5\textwidth]{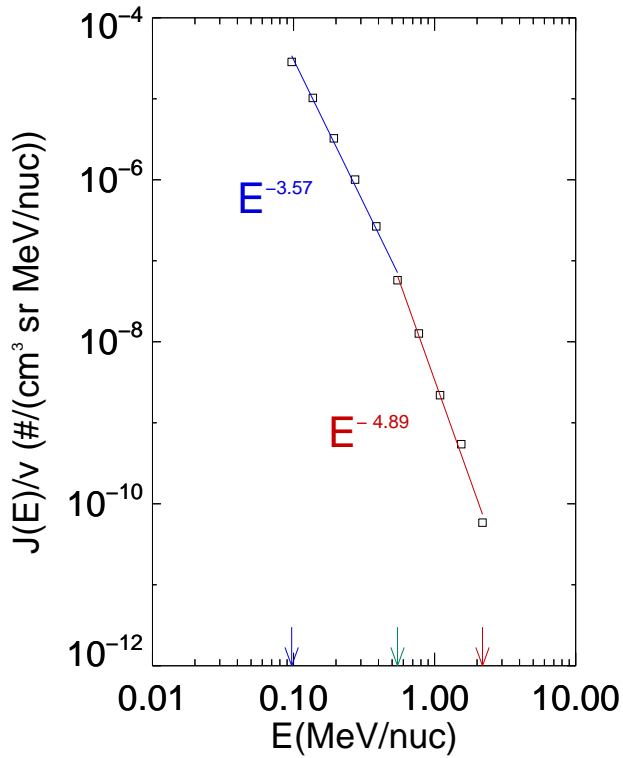}
\caption{Injection energy spectrum $f(E)$ obtained from Figure \ref{fig:sourceJ}. $f(E)=J(E)/v$, where $v$ is particle's speed. 
$f(E)$ is composed of two parts. The first part is $E^{-3.57}$ between $0.094$ MeV/nucleon $<E<$ $0.546$ MeV/nucleon and the 
second part is $E^{-4.89}$ between $0.546$ MeV/nucleon $<E<$ $2.185$ MeV/nucleon.}
\label{fig:source}
\end{figure}

\begin{figure}
\hspace{0.2cm}
\includegraphics[width=0.45\textwidth,angle=-90]{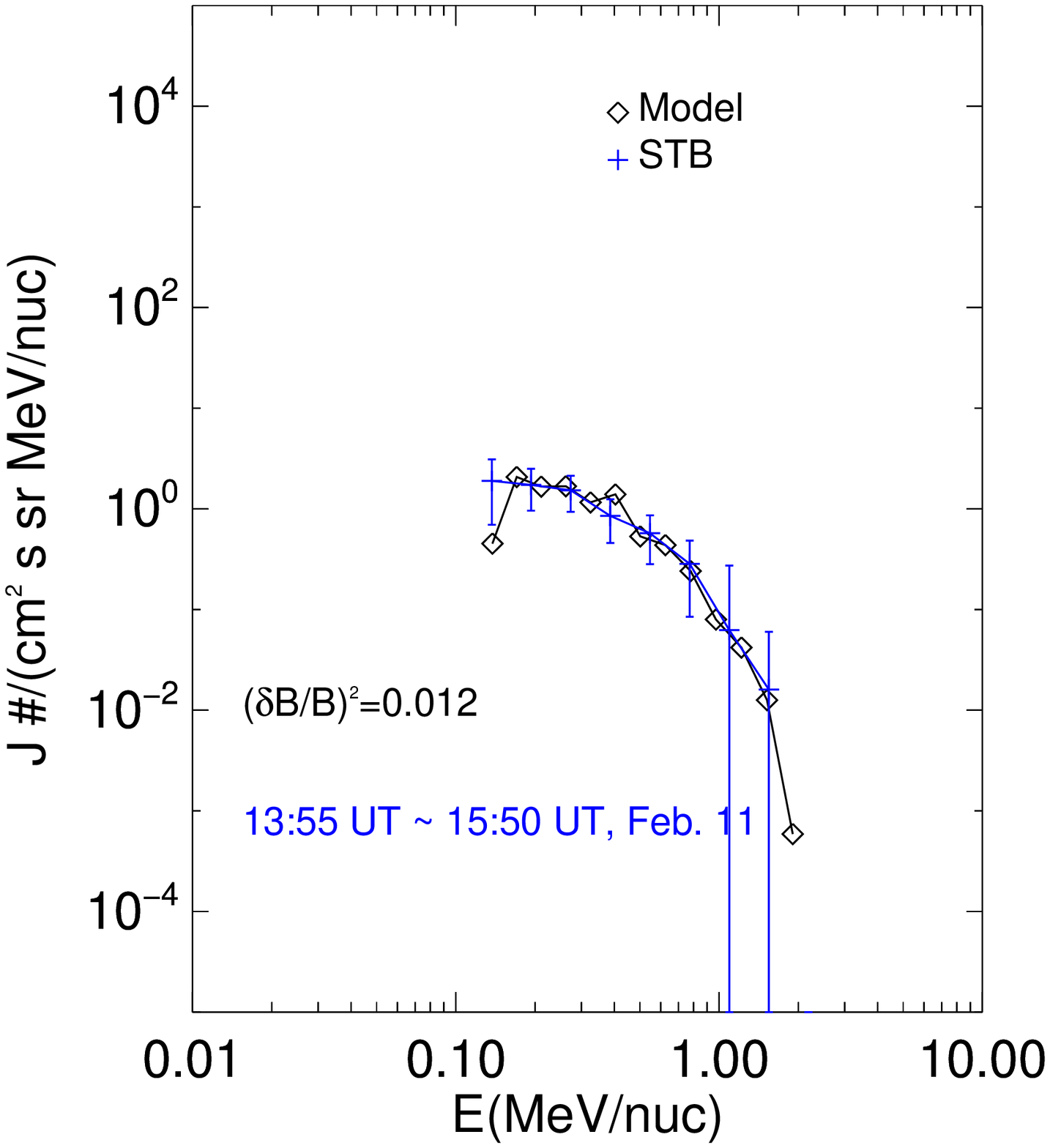}
\includegraphics[width=0.45\textwidth,angle=-90]{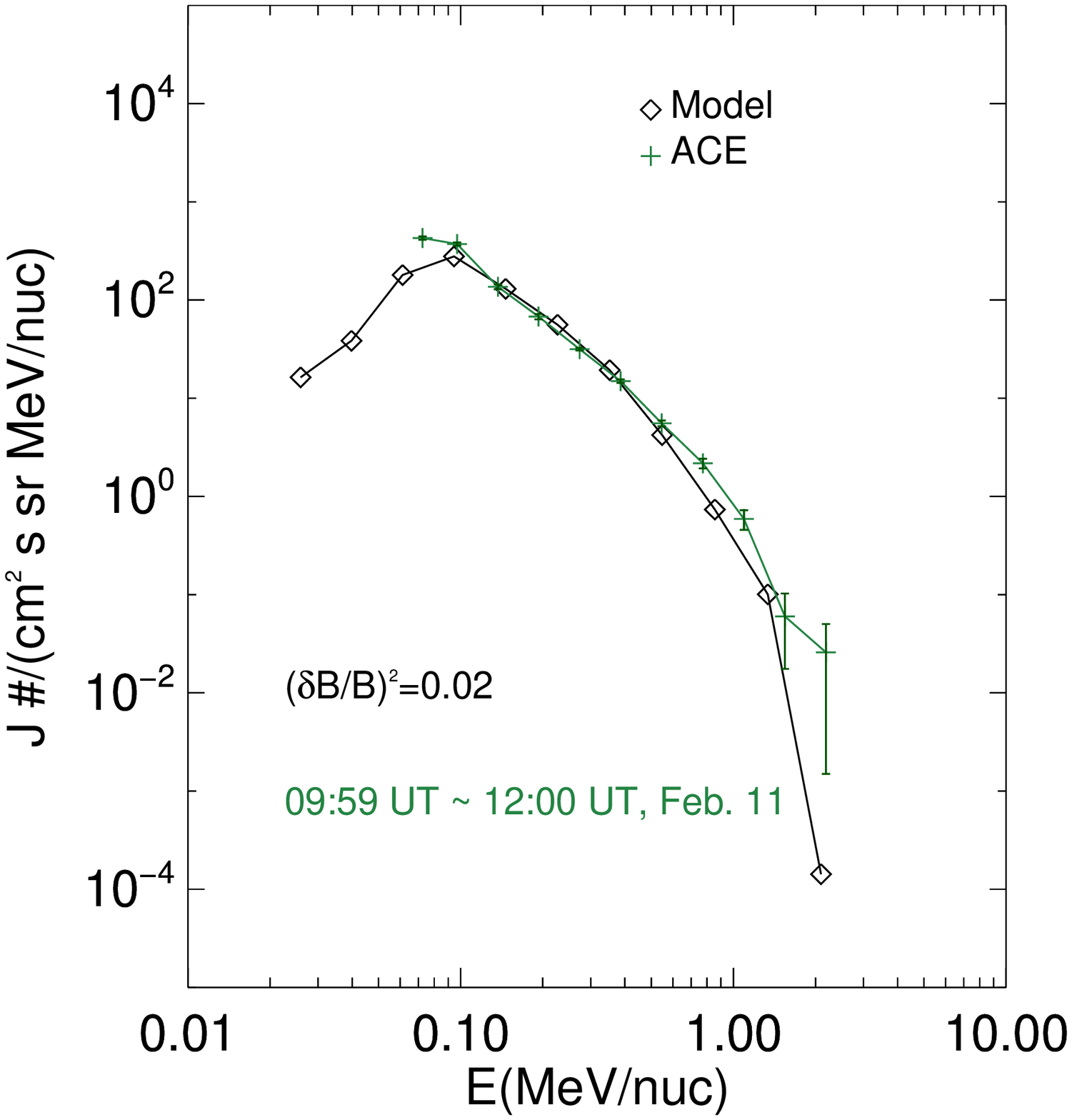}
\vspace{0.1cm}
\caption{Comparison of CIR-associated energetic particle differential intensities 
between observations and simulations. The left panel is for STB observation (case I) and 
the right pnael is for ACE observation (case II). 
The turbulence level at 1 AU for case I is $\delta^2 B/B_0^2=0.012$; and for case II is $\delta^2 B/B_0^2=0.02$.
Simulation results are shown in black and observations are shown in green and blue.}
\label{fig:result}
\end{figure}

\begin{figure}
\includegraphics[width=0.45\textwidth]{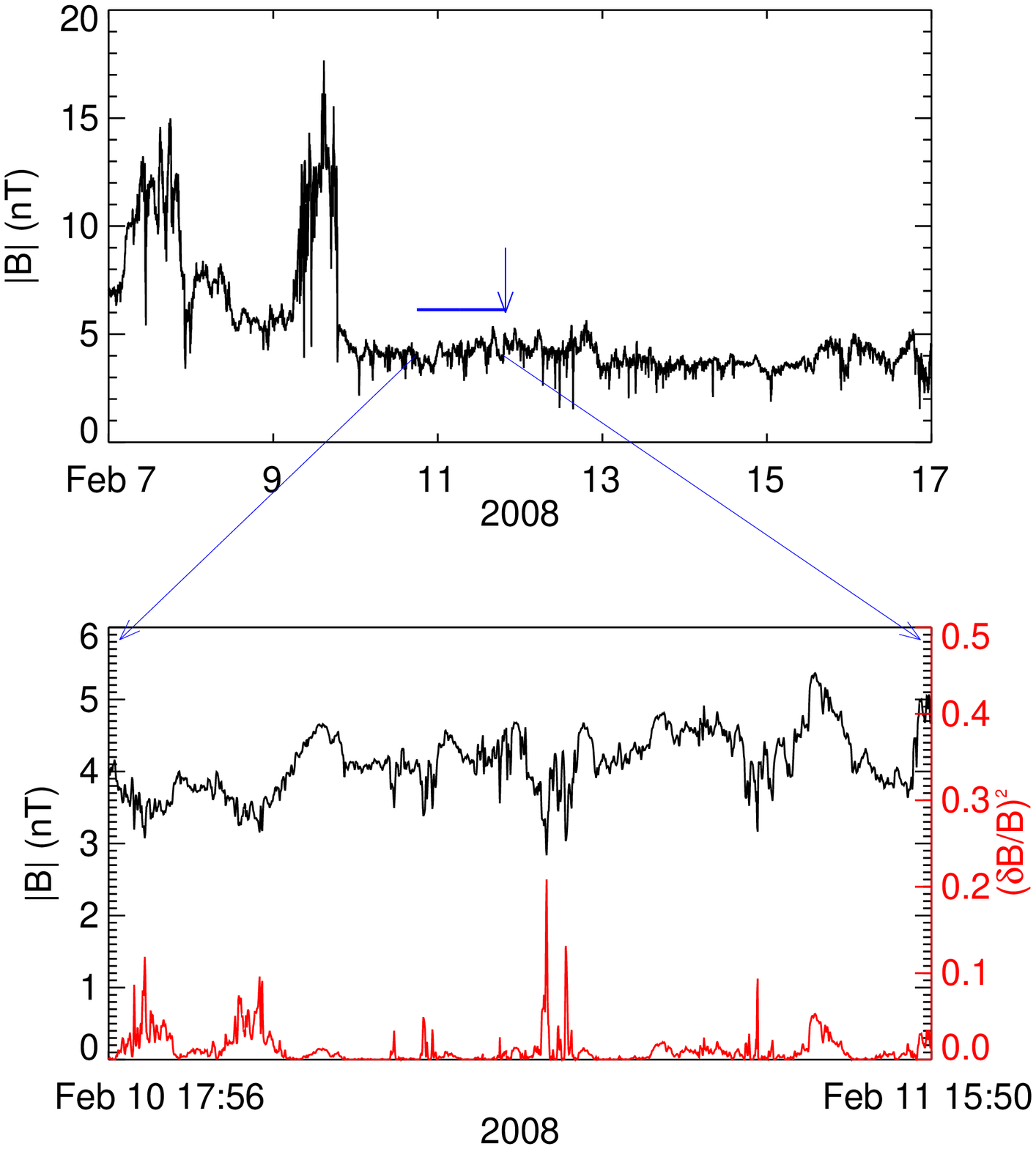} 
\includegraphics[width=0.45\textwidth]{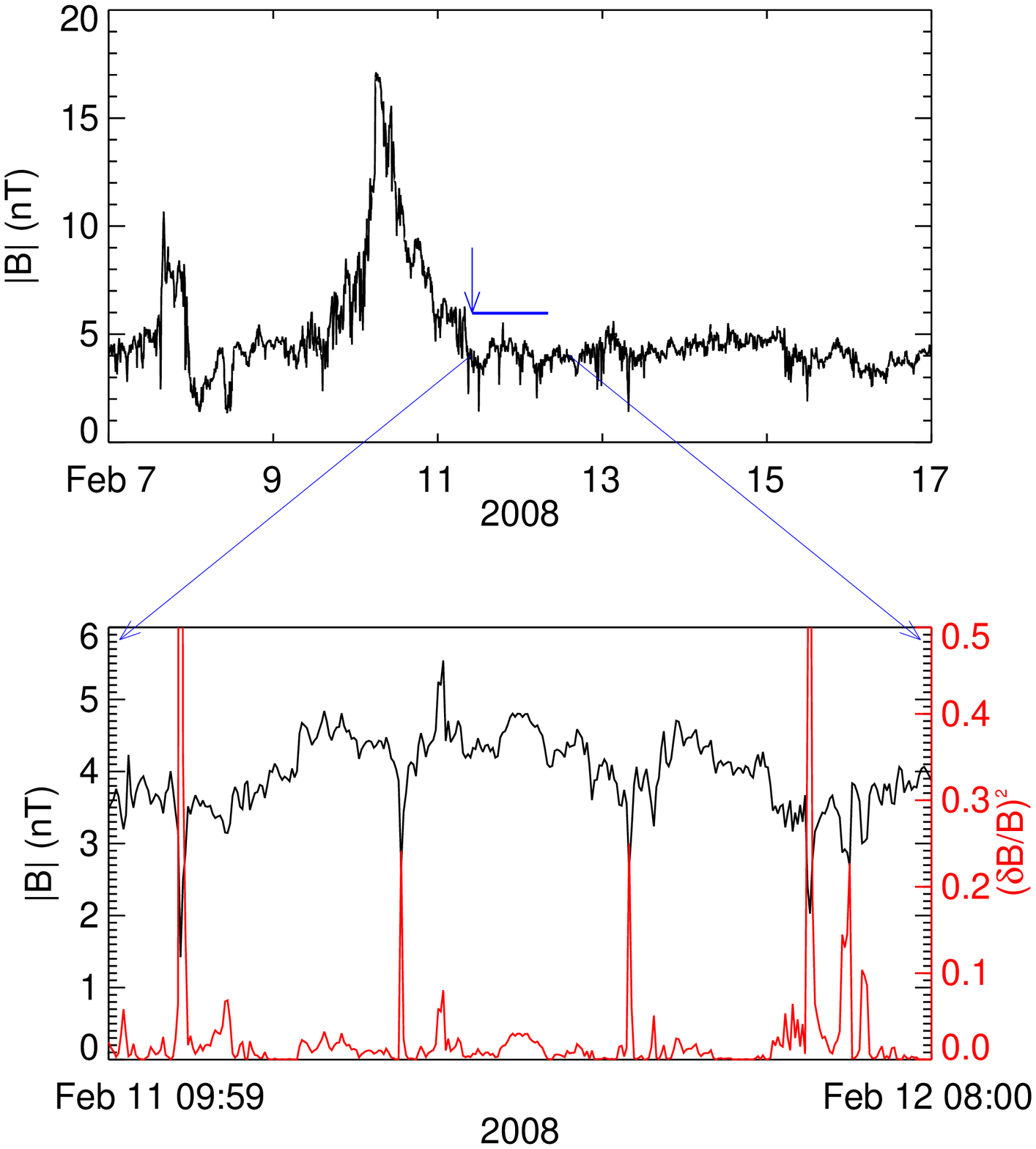} 
\caption{ STB (left) and ACE (right) observation of the magnetic field. 
 Upper left: observation of the magnetic field for the STB obsesrvation. 
 Upper right: observation of the magnetic field for the ACE obsesrvation. 
Lower left: A 20-hour zoom-in plot of the upper left panel. Red curve is $ (\delta B/B)^2$.
Lower right: A 20-hour zoom-in plot of the upper right panel. Red curve shows $ (\delta B/B)^2$. }
\label{fig:avgB}
\end{figure}

\begin{figure}
\includegraphics[width=1.0\textwidth]{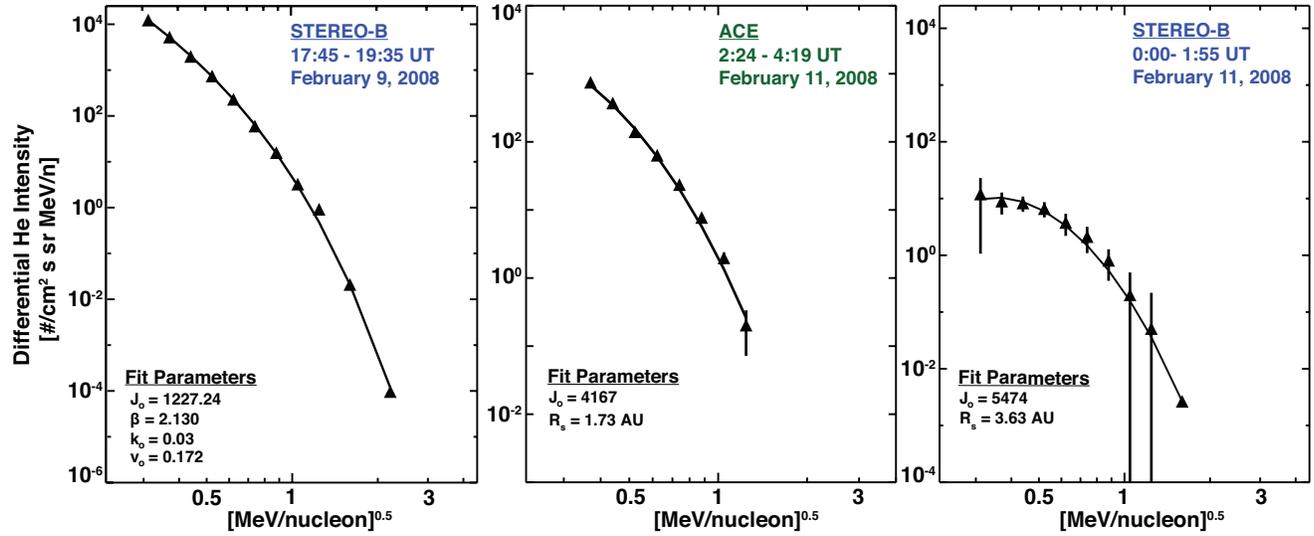}
\caption{The Fisk-Lee fitting of the differential intensities at, 1)  left panel:  the reverse shock as observed in-situ by STB 
during during $17:54$ UT $\sim 20:09$ UT, Feburary $09$, 2008 
2) middle panel:  upstream of the reverse shock for ACE observation 
during $02:30$ UT $\sim 04:30$ UT, Feburary $11$, 2008; 3) right panel:  upstream of the reverse shock for STB observation 
during  $00:00$ UT $\sim 02:0$ UT, Feburary $11$, 2008.}
\label{fig:FiskLeeFitting}
\end{figure}

\begin{figure}
\includegraphics[width=0.5\textwidth]{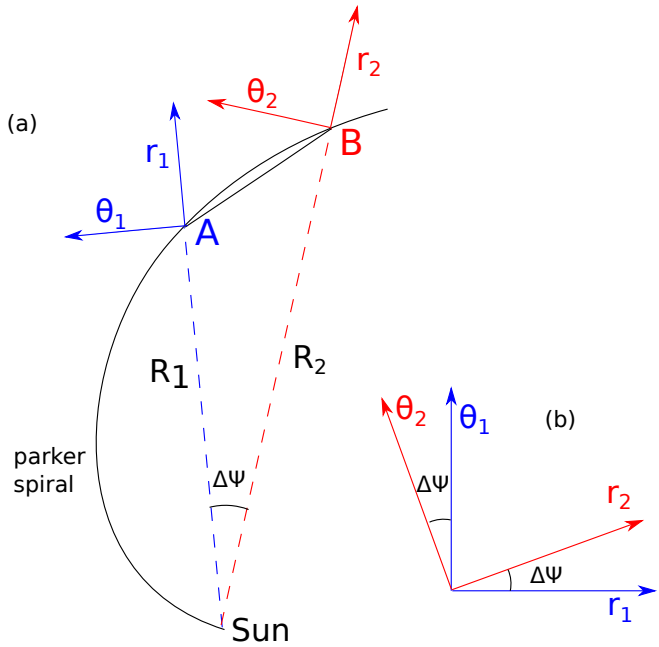}
\caption{Sketch of frame transformation. (a): $A$ and $B$ are two adjacent points in a Parker spiral. 
The radial distance of $A$ is $R_1$ and of $B$ is $R_2$. $\bm{r_1}$ ($\bm{r_2}$) is unit vector in the direction of 
$R_1$ ($R_2$).$\bm{\theta_1}$ ($\bm{\theta_2}$) is unit vector in the direction perpendicular to $\bm{r_1}$ ($\bm{r_2}$). 
$\Delta \psi$ is the angle between $\bm{r_1}$ and $\bm{r_2}$. (b): Relative directions of the four unit vectors.}
\label{fig:frametrans}
\end{figure}

\begin{figure}
\includegraphics[width=0.5\textwidth]{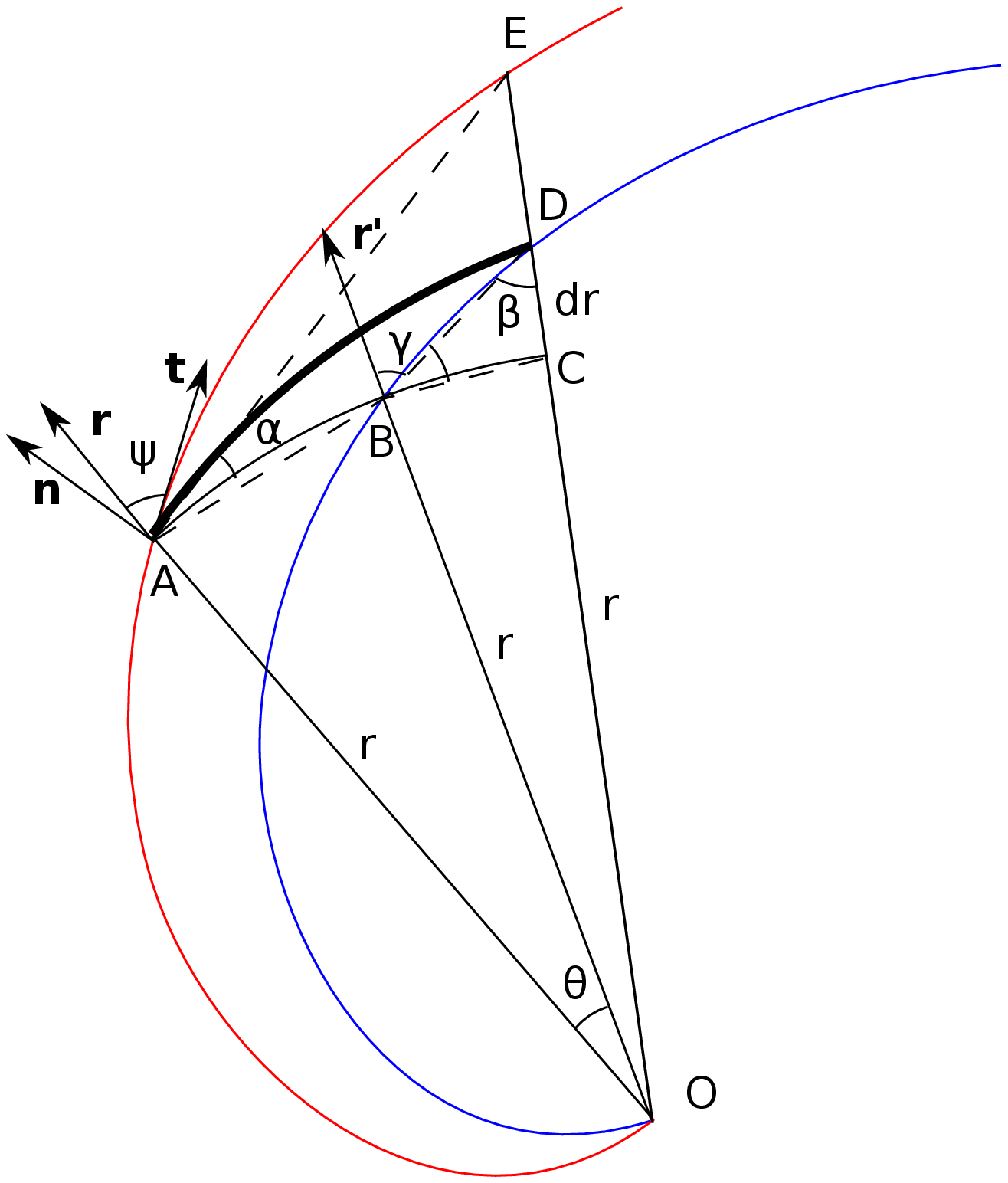}
\caption{CIR reverse shock configuration. The Sun is located at $O$. Red and blue curves are 
Parker's spiral. Black thick curve extending from $A$ to $D$ is reverse shock surface.}
\label{fig:inject}
\end{figure}

Total number of injected particles ($N_0$) at CIR reverse shock plays an important role in the simulation-observation comparison. Assuming the shock strength does vary along shock surface, the total number of injected particles is proportional to the seed particle density (n) and the shock cross section (A) in a flux tube. Figure \ref{fig:inject} shows the configuration of the reverse shock.  Similar to Figure \ref{fig:sketch}, red and blue curves are Parker spirals, black thick curve extending from $A$ to $D$ is the reverse shock surface.  The sun is located at $O$.  Red (blue) Parker spiral intersects the radial direction at $A$ and $E$ ($B$ and $D$).  The length $OA$ is the same as $OB$ and $OC$, which is $r$. The vectors $\bm{n}$, $\bm{r}$, $\bm{r'}$ and $\bm{t}$ are unit vectors  in the direction of shock normal, $OA$, $OB$ and tangent direction of the Parker spiral colored in red at point $A$ respectively. 

The following statements and calculations are based on the approximation that $\theta$, which is the angle between $OA$ and $OB$, is small.  Then, $AB$ is perpendicular to $OA$ and $OB$; $BC$  is perpendicular to $OB$ and $OD$; $AB$ and $BC$ are in the direction perpendicular to the radial direction.  Angle between $\bm{r}$ and $\bm{t}$ is $\psi$ and we have $tan\psi=B_t/B_r$ ($B_t$ and $B_r$ are given in Equation \ref{equ:spiralB}).  $\gamma$ is the angle between $\bm{r'}$ and $BD$, and $\beta$ is the angle between $BD$ and $CD$. And in the approximation, we have $\gamma=\psi$ and $\beta=\psi$. Angle between shock surface and $AB$ is $\alpha$, and the distance of $CD$ is $dr$.

Follow the law of cosine, in the triangle $ABD$, we get \begin{equation} AD^2+AB^2-2~AD~AB~\cos\alpha=BD^2. \label{equ:cosinelaw} \end{equation} The length of $AB$ is given by $r\theta$ and the length of $BD$ is given by $dr/\cos\beta$ (in the right triangle $BCD$), where $dr=AD~sin\alpha$ (in the right triangle $ACD$). We, therefore, get \begin{equation} AD=\frac{r\theta~\cos\alpha}{1-(\sin\alpha/\cos\psi)^2}(1+\tan\alpha~tan\psi). \label{equ:AD} \end{equation} 
In the triangle of $ACE$ and $ACD$, the relation between $tan\alpha$ and $tan\psi$ can be expressed as \begin{equation} \begin{split} tan\alpha&=\frac{CD}{AC}=\frac{V_{sh} dt}{AC},\\ cot\psi&=\frac{CE}{AC}=\frac{V_{sw} dt}{AC}, \end{split} \label{equ:tan} \end{equation} where $V_{sh}$ and $V_{sw}$ is the shock speed and fast solar wind speed and $dt$ is time for the sun to rotate by $\theta$ degree. Equation \ref{equ:tan} can be simplified into $tan\alpha~tan\psi=V_{sh}/V_{sw}$.

\end{article}

%
%

\begin{table}
\caption{Monte Carlo Results} \label{tab:MonteCarloFitting}
\centering
\begin{tabular}{lccccc}
\hline
  & {Shock Loc} & {$(\delta B/B)^2$} & {Spacecraft} & {$\alpha$} & {$\eta$} \\
\hline
Case I & 2.55 AU & 0.012 & STB & 6 & 1.27 \\
\hline
Case II & 1.39 AU & 0.02 & ACE & 1 & 0.61 \\
\hline
\end{tabular}
\end{table}

\begin{table}
\caption{Fisk \& Lee Fitting} \label{tab:FiskLeeFitting}
\centering
\begin{tabular}{lccccc} 
\hline
    &  {Shock Loc}  & {$\kappa_0$} & {Spacecraft} & {$J_0$ ( \# cm$^{-2}$ s$^{-2}$ sr$^{-1}$ (MeV/n)$^{-1}$) } & {$\eta$}  \\
\hline
Source & 1.0 AU & 0.03 & STB & 1227 & 1.0 \\
\hline
Case I & 3.63 AU & 0.03 & STB & 5474 & 3.4 \\
\hline
Case II & 1.73 AU & 0.03 & ACE & 4167 & 4.7 \\
\hline
\end{tabular}
\end{table}

\end{document}